\documentclass[twocolumn,epsfig,pre]{revtex4}
\usepackage{graphicx}
\usepackage{amssymb}
\usepackage{amsmath}
\usepackage{hyperref}
\usepackage{latexsym}

\def\be{\begin{equation}}
\def\ee{\end{equation}}
\def\bea{\begin{eqnarray}}
\def\eea{\end{eqnarray}}

\def\e{\epsilon}

\begin{document}
\title{A lattice gas model for active vesicle transport by molecular motors with opposite polarities}
\author{Sudipto Muhuri$^{1,2}$,  Ignacio Pagonabarraga$^{1}$}
\affiliation{$^1$ Departament de F\'{\i}sica Fonamental, Universitat de Barcelona, C. Mart\'{\i} i Franqu\'es 1, 08028-Barcelona, Spain\\
\noindent
$^2$Theoretische Physik, Universitat des Saarlandes- Postfach 151150, 66041 Saarbrucken, Germany}
\begin{abstract} 
We introduce a multi-species lattice gas model for motor protein driven collective cargo transport on cellular filaments. We use this model to describe and analyze the collective motion of interacting vesicle cargoes being carried by oppositely directed molecular motors, moving on a single biofilament. Building on a totally asymmetric exclusion process (TASEP) to characterize the motion of the interacting cargoes, we allow for mass exchange with the environment, input and output at filament boundaries and focus on the role of interconversion rates and how they affect the directionality of the net cargo transport. We quantify the effect of the various different competing processes in terms of non-equilibrium phase diagrams. The interplay of interconversion rates, which allow for flux reversal and evaporation/deposition processes introduce qualitatively new features in the phase diagrams. We observe regimes of three-phase coexistence, the possibility of phase re-entrance and a significant flexibility in how the different phase boundaries shift in response to changes in control parameters. The moving steady state solutions of this model allows for different possibilities for the spatial distribution of cargo vesicles, ranging from homogeneous distribution of vesicles to polarized distributions, characterized by inhomogeneities or {\it shocks}. Current reversals due to internal regulation emerge naturally within the framework of this model. We believe this minimal model will clarify the understanding of many  features of collective vesicle transport, apart from serving as the basis for building more exact quantitative models for vesicle transport relevant to various {\it in-vivo} situations.  

\end{abstract}

\maketitle
\section{Introduction}
Biofilaments such as microtubules and actin play a central role in ensuring the mechanical integrity of eukaryotic cells~\cite{cell} and intracellular transport. The entire network of actin and microtubule filaments serve as tracks for motor protein assisted transport of cargo vesicles and organelles from one specific location to another~\cite{cell,howard}. This transport process is active: multiple motor proteins bind to the filaments, and use ATP hydrolysis to convert the stored chemical energy into mechanical work ~\cite{howard, welte}. In fact the heterogeneous distribution of organelles and vesicles and their regulation within the cell is achieved through this motor protein driven active transport process ~\cite{welte, bray}. The lack of detailed balance which underlines the active motion of the molecular  motors on filaments, renders the collective behavior of such a system  markedly different from the one for an assembly of passive molecules. 

Biofilament polarity provides a natural means to direct molecular motors, and depending on their molecular structure, different motor families  displace to one of the two end of the  filaments on consumption of ATP~\cite{howard}.  For example, for microtubules (MT),  dynein motors move toward the MT minus end while kinesin is a plus end directed motor~\cite{gross}. The existence of motors moving in opposite directions allows for bidirectional cargo transport.  {\it In-vitro} experiments  using fluorescence video microscopy with melanophore  cellular extracts of cellular filaments( actin and microtubules), motor proteins and vesicle  cargoes~\cite{borisy1, borisy2} has shown that under ATP rich conditions,  individual cargo vesicles  move bidirectionally on MT in the presence of oppositely moving motors such as dynein and kinesin~\cite{gelfand}.

A single cargo vesicle will be usually attached to  several molecular motors~\cite{gelfand, lipo_bead, lipo1}. If these motors have opposing polarity, the net cargo displacement along the filament will arise as a result  of motor competition.  Under these circumstances, interconversion or directional switching arising from  motor competition and their kinetics to the filament track is required to ensure efficient  cargo vesicle transport over long distances  and avoid jamming.  The origin of this competition, and the existence of regulatory mechanisms  which determine the collective behavior of such motors remains an object of controversy~\cite{gross_06}. While specific regulatory molecules have been proposed~\cite{welte} recent theoretical studies have shown that regulation can arise as a result of the non-linear dependence of the motor velocity to the forces it is subject to~\cite{lipo_tugofwar}. Therefore, the traditional clear distinction between   tug-of-war~\cite{gelfand}, wher
Even if insight has been gained in the molecular  mechanisms underlying the transport of a single vesicle, much less is understood regarding  bidirectional collective motion of vesicles~\cite{prost_jullicher}. Experimental evidence has shown that the spatial distribution of vesicles is sensitive to the cellular environment. Hence, their dispersion does not  depend only on  specific biochemical signals, which alter the motor binding properties, but can be controlled by 
modifying the polymerization degree of the embedding actin  network. In experiments carried out with cellular extracts of fish melanophores, increasing the amount of caffeine, a chemical which alters motor binding kinetics,  leads to a uniform dispersion of vesicle granules. On the contrary, adding latrunculin, which favors actin depolymerization, the granules loose their ability to disperse and accumulate in the cell periphery~\cite{borisy1, welte}.   At high cargo concentration, of relevance in {\it in vivo} situations, one cannot ignore inter-vesicle interactions. Studies carried out on  mitochondrial distribution and movement in cultured neurons highlights the importance of inter-organelle interaction~\cite{hollenbeck}. Moreover,  transport is affected by cargo loading and removal at the filament end.  For instance a sink mechanism has been proposed for transport of pigment granules in mammalian melanocytes, wherein granules are captured by myosin V on reaching the filament end~\cite{welte,boundaryref}.

These facts highlight the relevance of the interplay of different processes determining vesicle  distribution. Coarse grained models  which incorporate these competing  processes become useful means to rationalize their effects at large scales and describe the collective features resulting from bidirectionality. We will  make use of such an approach to analyze quantitatively  the emergence of spatial organization and regulation of vesicle distribution induced by active motor transport.  We will propose  an  effective model   based on the  totally asymmetric exclusion process (TASEP)~\cite {tasep1,tasep2,tasepexact,kolo}. TASEP exhibits non-equilibrium phase transitions between different macroscopic density and current states. Variants of this model have been studied  to model  collective unidirectional motion of molecular motors or other cargoes, whenever the detailed motor molecular interactions can be disregarded ~\cite{lipo2,tasepsugden, freylet, explipo}. Despite its simplicity, this theoretical approach predicts  inhomogeneous motor density profiles and  shock development  along the filament~\cite{lipo2} and provides a simple reference framework to address generic issues related to collective motor transport~\cite{houtman}. TASEP has also been used to address the effective interaction  with the filament environment,  either because of  diffusion or active transport through  nearby filaments, through  particle attachment/detachment, analogous to a Langmuir kinetic process (LK). The phase diagram of  TASEP-LK has shown the persistence of density shocks and the coexistence of different motor phases which characterize their collective transport  behavior~\cite{santen,freypre}.   

Although, in general, a single vesicle cargo can be attached to a variable number of  molecular motor, Ref.~\cite{lipo_tugofwar} has shown that the cumulative effect of competing motors  can effectively be described in terms of  right-  and left-moving species and that adding a resting species improves the  comparison with experimental results on single moving  vesicles~\cite{explipo}.  For the sake of simplicity, we introduce a generalized two-species TASEP-LK model to describe bidirectional collective cargo transport. Including a third, non-moving species, is straightforward  and does not alter the qualitative conclusions we will discuss subsequently. The minimal model that we first put forward in Ref.~\cite{igna}, accounts for  the non-conservative and collective transport of cargo vesicles on a biofilament, which includes the effects of excluded volume interaction between the cargo vesicles, evaporation-deposition processes of the vesicles in the bulk and boundary loading/off-loading of cargo vesicles at the boundaries. We will concentrate in a simple open geometry where vesicles enter from the extreme of the filament, as opposed to cyclic geometries, where for a single motor species a homogeneous motor distribution is expected.  We will combine mean field analytical studies with Monte Carlo simulations to characterize the role of the boundaries in the  structure and transport inside the filament. We will discuss when shocks develop and will determine the corresponding  non-equilibrium phase diagrams, which show boundary and motor  current reversals along with domain wall(shocks) localization along the filament.
 This simple theoretical framework allows us  to identify the relevant physical processes which control the appearance and stability of the  collective behavior of  coexisting motors with opposed polarity. Hence, despite its simplicity, the proposed framework  will  provide qualitatively understanding of the relevant  features of bidirectional vesicle transport on microtubule networks.  In Section~\ref{sec:model} we introduce the model we will study and  in Section~\ref{sec:mf} we derive the mean field (MF) continuum equations for the vesicle densities and describe also the numerical scheme we have implemented to check the MF theoretical predictions.  Due to the open geometry under scrutiny,  we analyze in detail the boundary conditions and their continuum limit. In Section~\ref{sec:boundary} we obtain and analyze the density and current profiles solving the MF equations obtained in the previous section. In Section~\ref{sec:phase} we construct the phase diagram after having obtained the corresponding phase boundaries. We follow it up with a discussion on the nature of the phase diagrams and conclude in Section~\ref{sec:disc} with a discussion of the main results and their implications for cargo vesicle transport. 
\section{The minimal model}
\label{sec:model}

\begin{figure}
\begin{center}
\includegraphics [width=3.0in,angle=-90]{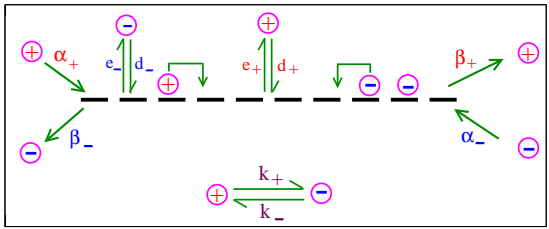}
\end{center}
\caption{(Color online) Schematic representation of all the dynamical processes for right (+) and left(-)-moving cargoes. The left and the right moving cargoes hop to the neighbouring site with rate 1 if the site is empty. The rates of entry, $\alpha_{\pm}$,  and exit, $\beta_{\pm}$, of cargoes at the filament boundaries are displayed along with the adsorption, $d_{\pm}$, and desorption, $e_{\pm}$, rates for the left and right moving cargoes. At each node in the lattice, the interconversion rates from right (left) moving state to left (right) moving one are characterized by the rates $k_{\pm}$.}
\label{delta01}
\end{figure}
We represent the biofilament as a finite one-dimensional lattice of length $L$ with $N$ sites, labeled ~$i = 0, \dots,N-1$ and the  lattice spacing, $\epsilon = L/N$,  is related  the basic cargo vesicle displacement. The filament ends are open and control the net cargo flux with the environment through specific kinetic rules.

 We characterize the microscopic state of the system at a given site $i$ at time $t$ in terms of the occupation numbers of the cargoes moving to the right,~ $ n_{i}^{+}$, and to the left, $n_{i}^{-}$. For microtubules they  would correspond  to  the motion of cargo vesicles induced by kinesins and dynein motors, respectively. The cargo size prevents multiple occupancy, irrespective of the directionality of motion. Thus only  one particle can occupy a given node at a given time so that the occupation numbers take values 0 and 1. The filament voids occupation, $~n_{i}^{0}$, can be regarded as  a useful auxiliary species, determined by the conservation law, $ n_{i}^{+} + n_{i}^{-} + n_{i}^{0}= 1$.\\

For the sites $i=0, \ldots, N-2$, right-moving particles (+) can hop to the neighbouring site at $i + 1$ if it is empty  with a rate $k_t$. Correspondingly, for the sites $ i=1,\ldots, N-1$, left-moving cargoes $(-)$ can hop to the neighbouring site at $i - 1$ if the site is empty  with a rate $k_t$; subsequently  the inverse of this  hopping rate will be taken as the unit of time.

At the  lattice ends, vesicle cargoes can enter and/or leave depending on their polarity. Therefore, at site $i = 0$, right-moving particles (+) can enter the filament with rate $\alpha_{+}$, if the site is empty and left moving particles can leave it with a rate $\beta_-$. At the opposite end, on site $i = N-1$, (+)-particles leave the lattice with rate, $\beta_{+}$ and (-)-cargoes enter with rate $\alpha_-$ if the site is empty. 

Due to the finite processivity of the molecular motors, cargoes can detach from the filament, and diffusing vesicles can also approach and attach. The details of this interaction, however, depends on the  properties of the medium. For example, a surrounding actin network  allows for the jump of vesicles to the neighbouring tracks and the attachment is controlled by such active transport processes. We include  vesicles hopping off and on the filament as  kinetic attachment and detachment processes. For all sites in the bulk, $i = 1, N-2 $, $(+)$ and $(-)$ moving cargoes  can detach  from the filament with site-independent rates    $e_+$  and  $e_-$ and  attach at an unoccupied node  with rates $d_+$ and $d_-$  respectively.
Finally, the interaction between motors of opposite polarity is accounted for by a local interconversion process between the two opposing  species. We characterize such a process  by an interconversion rates, $k_+$, $(k_{-})$, corresponding to the rate at which a right(left) moving cargo becomes a left(right) moving one. These rates are considered uniform along the filament. We display the relevant dynamical processes in Fig.~\ref{delta01}.

\subsection{Relevant time scales and dynamic regimes}

Since there are different competing dynamical processes, qualitatively different dynamical regimes  can be identified depending on the rate limiting mechanism.  If all the rates exhibit the same scaling with  particle number, for a sufficiently long filament the dynamics corresponds to that prescribed by the attachment/detachment processes, and the behavior is characteristic of Langmuir dynamics, except for a narrow region near the filament ends~\cite{freylet}. This is the  situation if  processes are determined only  by local  interactions at the nodes because the mean number of visited sites along the filament before a cargo detaches  is of order $1/e_{\pm}$. For  attachment/detachment rates which scale extensively, a new dynamical regime is achieved where incoming fluxes compete with transport and desorption processes. An alternative regime,where the  environment exchange rates are dominant has also  been explored in  a related  model for  bidirectional motors~\cite{KLUMPP_EPL}, which constitutes a generalized TASEP-LK scenario. We will concentrate in {\bf the regime for which the } evaporation and adsorption rates scale {\bf inversely} with the filament size; hence we introduce the rescaled rates $E_{\pm}=e_{\pm}N$ and  $D_{\pm}=d_{\pm}N$~\cite{freylet}.\\
We will restrict ourselves to the scenario where interconversion is the fastest process, relevant for bidirectional vesicular transport where  opposite motors appear to engage in rapid succession~\cite{gross_06}. In this limit, at each site $i$, the density of $(+)$ species equilibrates with the density of $(-)$ particle and  the two species become proportional to each other, as determined by the ratio of interconversion rates. We will also see that in this regime  the ratio of interconversion rates will determine the  overall direction of the  cargo flux.  Moreover, the existence of two competing species breaks the particle/hole symmetry  characteristic of  one-species TASEP models, through the interconversion process. This lack of symmetry will affect significantly the non-equilibrium phase diagram of the system.

\subsection{Equations of motion}
Once we have established the dynamics of the two  cargo species, we can write down the corresponding  evolution equations for their occupation numbers. Assuming that the displacement rate is fast enough, we can treat time as a continuous variable, leading to
\begin{eqnarray}
 \frac{dn_i^{+}}{dt} &=& n_{~i-1}^{+}(1 - n_{i}^{+} -n_{i}^{-}) - n_{i}^{+}(1 - n_{~i+1}^{+} -n_{~i+1}^{-})\nonumber\\
&-&~k_{1}n_{i}^{+}+ k_{2}n_{i}^{~-}-e_{+}n_{i}^{+} + d_{+}(1 - n_{i}^{+} - n_{i}^{~-})\nonumber \\
\frac{dn_i^{-}}{dt} &=& n_{~i+1}^{-}(1 - n_{i}^{+} -n_{i}^{-}) - n_{i}^{-}(1 - n_{~i-1}^{+} -n_{~i-1}^{-})\nonumber\\&-&~k_{2}n_{i}^{-}+ k_{1}n_{i}^{+}-e_{-}n_{i}^{-} + d_{-}(1 - n_{i}^{+} - n_{i}^{~-}) 
\label{eq:rate_1}
\end{eqnarray}
\indent
The terms on the r.h.s of Eq.(\ref{eq:rate_1})  describe  gain and loss processes at a given node $i$  arising from displacement, inter-conversion and deposition-evaporation events. Associated with displacement, we can identify the particle flux for the (+)- and (-)- particles at node $i$
\begin{equation}
J^{\pm}_{i} = n_{i}^{\pm}(1 - n_{~i\pm1}^{+} -n_{~i\pm1}^{-}).
\end{equation}

The boundary conditions are readily expressed in terms of the instantaneous vacancy number, $n_{i}^{0}$.
At the left boundary site, $i = 0$, a vacancy enters, when either a $(-)$ particle leaves the left boundary or a $(+)$ particle hops to the neighbouring site in the bulk. Similarly a vacancy leaves the boundary site at the left boundary when either a $(+)$ enters the boundary site, $i = 0$ or a $(-)$ particle enters this site the next neighbouring site in the bulk, $i = 1$ Accordingly,
\begin{equation}
\frac{dn_0^{0}}{dt}  = n_{0}^{+} n_1^0 + \beta_{-}n_{0}^{-} -\alpha_{+}n_0^0-n_{1}^{-} n_0^0
\label{eq:n0}
\end{equation}
 The terms on the right hand side of the equation above have a simple interpretation of gain and loss terms, with the first two terms contributing to the entry rate of vacancies, while the last two terms correspond to the exit rate of vacancies at the boundary site. 
Similarly, the evolution of instantaneous vacancies at the right  filament boundary reads
\begin{equation} 
\frac{dn_{N-1}^{0}}{dt} = n_{N-1}^{-} n_{N-2}^{0} + \beta_{+}n_{N-1}^{+} -\alpha_{-} n_{N-1}^{0}-  n_{N-2}^{+} n_{N-1}^{0}
\label{eq:n1}
\end{equation}

Averaging the occupation numbers  over  a time interval $dt$ and over initial conditions,  the corresponding equations of motion for the expectation values are given by,

\begin{eqnarray}
\frac{d\langle n_i^{+}\rangle}{dt} &=& \langle n_{i-1}^{+}(1 - n_{i}^{+} -n_{i}^{-})\rangle -\langle n_{i}^{+}(1 - n_{i+1}^{+} -n_{i+1}^{-})\rangle \nonumber\\
&-&~k_{+}\langle~n_{i}^{+}\rangle + k_{-}\langle~n_{i}^{-}\rangle -e_{+}\langle~n_{i}^{+}\rangle \nonumber\\&+& d_{+}\langle~1 -n_{~i+1}^{+} -n_{~i+1}^{-}\rangle \nonumber\\
\frac{d\langle n_i^{-}\rangle}{dt} &=& \langle n_{i+1}^{-}(1 - n_{i}^{+} -n_{i}^{-})\rangle - \langle n_{i}^{-}(1 - n_{i-1}^{+} -n_{i-1}^{-})\rangle \nonumber\\&-&~k_{-}\langle~n_{i}^{-}\rangle + k_{+}\langle~n_{i}^{+}\rangle-e_{-}\langle~n_{i}^{-}\rangle \nonumber\\&+& d_{-}\langle~1 - n_{~i+1}^{+} -n_{~i+1}^{-}\rangle 
\label{eq:average}
\end{eqnarray}
and accordingly, the mean void density at the ends of the filament can be expressed as
\begin{eqnarray}
\frac{d\langle n_0^{0}\rangle}{dt}  &=& \langle n_{0}^{+} n_1^0\rangle + \beta_{-}\langle n_{0}^{-}\rangle -\alpha_{+}\langle n_0^0\rangle-\langle n_{1}^{-} n_0^0\rangle\nonumber\\
\frac{d\langle n_{N-1}^{0}\rangle}{dt} &=& \langle n_{N-1}^{-} n_{N-2}^{0}\rangle + \beta_{+}\langle n_{N-1}^{+}\rangle -\alpha_{-} \langle n_{N-1}^{0}\rangle\nonumber\\&-&  \langle n_{N-2}^{+} n_{N-1}^{0}\rangle
\label{eq:average_1}
\end{eqnarray}
The above set of equations are exact and, in principle, the steady states or the time evolution of the expectation values can be determined. However knowing the expectation value of any quantity requires the knowledge of all higher order correlations
which makes it intractable analytically~\cite{privman}. In specific cases, like TASEP, this moment hierarchy can be handled exactly using matrix method approach ~\cite{tasepexact}. Alternatively, if the number density fluctuations are smaller than mean density, a mean field approach works remarkably well~\cite{kolo}

\section{Mean Field(MF), Continuum limit and Monte Carlo (MC) simulations}
\label{sec:mf}

\subsection{Mean Field and Continuum limit}
We define the mean occupation densities as  $\rho_i^{+} \equiv \langle n_{i}^{+}\rangle$ and $\rho_i^{-} \equiv \langle n_{i}^{-}\rangle$, and we will introduce a mean field theory in which  we factorize   all the two-point correlators arising out of the different combinations of $n_{i}^{+}, n_{i}^{-}$ ~as a product of their averages, 
\begin{eqnarray}
\langle n_{i}^{\pm}n_{i+1}^{\pm}\rangle &=& \langle n_{i}^{\pm}\rangle \langle n_{i+1}^{\pm}\rangle = \rho_i^{\pm}\rho_{i+1}^{\pm}\nonumber\\
\langle n_{i}^{\pm}n_{i+1}^{\mp}\rangle &=& \langle n_{i}^{\pm}\rangle \langle n_{i+1}^{\mp}\rangle = \rho_i^{\pm}\rho_{i+1}^{\mp}\nonumber 
\end{eqnarray}
where the  corresponding expression for the currents can be expressed as,
\begin{equation}
J^{\pm}_{i} = \langle n_{i}^{\pm}\rangle \langle(1 - n_{~i\pm1}^{+} -n_{~i\pm1}^{-})\rangle =\rho_i^{\pm} (1-\rho_{i\pm 1}^+-\rho_{i\pm i}^-)
\end{equation}
If the filament length, $L$, is long compared with the cargo size, it is reasonable to keep $L$ fixed and increase the number of lattice sites,  $N\rightarrow \infty$, so that the lattice spacing $\epsilon = \frac{L}{N}\rightarrow 0$~\cite{freylet} and the position along the filament can be expressed as $x = \frac{i}{N-1}$, with $0 \leq x \leq 1$. In this continuum limit, the mean densities  can be written as
\begin{eqnarray}
\langle~n_{i}^{+}~\rangle &=& \rho_{+}(x, t)\nonumber \\
\langle~n_{i\pm1}^{+}~\rangle &=& \rho_{+}(x, t) \pm \epsilon\frac{\partial}{\partial x}\rho_{+}(x, t) + \frac{\epsilon^{2}}{2}\frac{\partial^2}{\partial x^2} \rho_{+}(x, t)
\end{eqnarray}
up to second order in $\epsilon$. Accordingly, the discrete mean field evolution equation, Eq.~(\ref{eq:average}), with the corresponding boundary conditions, Eqs.~(\ref{eq:average_1}), reduce to
\begin{eqnarray}
 \frac{\partial \rho_{+}}{\partial t}&=& - k_{+}\rho_{+} + k_{-}\rho_{-} + \e D_{+}(1 - \rho_{+} -\rho_{-}) - \e E_{+}\rho_{+} \nonumber\\
&& -\e \frac{\partial}{\partial x}\left[\rho_{+}(1 - \rho_{+} -\rho_{-})\right] + \frac{\e^{2}}{2}\left[\frac{\partial^{2}\rho_{+}}{\partial x^2}(1 - \rho_{+} -\rho_{-})\right]\nonumber\\ && + \frac{\e^{2}}{2}\left[(\frac{\partial^{2}\rho_{+}}{\partial x^2} +\frac{\partial^{2}\rho_{-}}{\partial x^2})\right]\rho_{+}
\label{eq:densityeqn1}
\end{eqnarray}
\begin{eqnarray}
 \frac{\partial \rho_{-}}{\partial t}&=& - k_{-}\rho_{-} + k_{+}\rho_{+} + \e D_{-}(1 - \rho_{+} -\rho_{-}) - \e E_{-}\rho_{-} \nonumber\\
&& +\e \frac{\partial}{\partial x}\left[\rho_{-}(1 - \rho_{+} -\rho_{-})\right] + \frac{\e^{2}}{2}\left[\frac{\partial^{2}\rho_{-}}{\partial x^2}(1 - \rho_{+} -\rho_{-})\right]\nonumber\\&+& \frac{\e^{2}}{2}\left[(\frac{\partial^{2}\rho_{+}}{\partial x^2} +\frac{\partial^{2}\rho_{-}}{\partial x^2})\right]\rho_{-}
\label{eq:densityeqn}
\end{eqnarray}
Rather than working with the two species densities, it will be useful to re-express Eq.~(\ref{eq:densityeqn1}) and Eq.~(\ref{eq:densityeqn}) in terms of the total cargo density and the relative concentration of the different species,
\begin{equation}
\rho = \rho_{+} + \rho_{-}, \;\;\;\;\;\;\;\; \phi = \rho_{+} - \rho_{-}
\end{equation}
which yield  $ 2\rho_{\pm} = \rho \pm \phi$. Accordingly, we can  rewrite the evolution equations to obtain
\begin{eqnarray}
 \frac{\partial \rho}{\partial t }&=& \e D_0(1 - \rho)\nonumber\\ &-&\frac{\e}{2}\left[E_{+}(\rho + \phi) + E_{-}(\rho - \phi)\right] -\e\frac{\partial}{\partial x}\left[\phi(1 - \rho)\right]  \label{eq:mf_rho}\\
\frac{\partial \phi}{\partial t}&=& -k_{+}(\rho + \phi) + -k_{-}(\rho- \phi) + \e(D_{+}- D_{-})(1 - \rho)\nonumber\\ &-& \frac{\e}{2}\left[E_{+}(\rho + \phi) - E_{-}(\rho - \phi)\right] -\e\frac{\partial}{\partial x}\left[\rho(1 - \rho)\right] 
\label{eq:mf_phirho}
\end{eqnarray}
where we have introduced $D_0=D_{+}+ D_{-}$, and have retained terms to leading order  in $\epsilon$. 
These equations show that the dependence in terms of relative concentration is linear, and that non-linearities are only related to  the global density occupation; the density dependence shows similarities with  TASEP for a single species.

We are interested in the steady state properties. Accordingly, we set  $\frac{\partial \rho}{\partial t} = \frac{\partial \phi}{\partial t} = 0 $. Since the interconversion rate is the fastest process, in the limit of small $\epsilon$, to leading order the two densities become proportional to each other,
\begin{equation}
\phi = \frac{(k_{-} - k_{+})\rho}{(k_{+} + k_{-})}=\frac{1-K_+}{W_+}\rho(x)
\label{eq:ss_1}
\end{equation}
where we  have introduced the effective ratio  of interconversion rates   $K_{+} = \frac{k_{+}}{k_{-}}$, which  determines the amount of species  imbalance at each node, and $W_+=1+K_+$.   From this relation, it is straightforward to obtain the corresponding  relation between the local occupancy of each species   
\begin{equation}
\rho_{+}(x) = K_{+}\rho_{-}(x)
\label{eq:ratio}
\end{equation}
which shows that   $K_+$, controls the local fraction of each species on the biofilament. For equal interconversion rates, the occupation is symmetric. 

The expression for current of $(+)$ species reads in general 
\begin{equation}
J_{+}(x)= \rho_{+}(x)\left[1 - W_{+}\rho_{+}(x)\right]\;\;\;\;\;
\end{equation}

Using Eq.~(\ref{eq:ratio}) it follows that the fluxes   of left- and right-moving cargoes are proportional to each other,  $J_{-} = -K_{+}J_{+}$. The total particle current along the filament reduces to,  
\begin{equation}
J(x) = J_{+}(x) + J_{-}(x)=(1-K_+)J_+(x)
\end{equation}
which explicitly shows that motor regulation allows for current reversal.   For  $0 < K_{+} < 1$, the overall flux of vesicles is from left to right while for $K_{+} > 1$, the overall flux is from right to left. 
Thus we see that for a fixed set of incoming  and outgoing particle fluxes, a change in the  motor expression can lead to total cargo flux  reversal.
Different motor types exhibit different relative  binding affinities~\cite{gelfand}. Accordingly, symmetric motor interconversion, $K_+=1$,  when  there is no net motor flux and cargo kinetics  reduces to that of  symmetric exclusion process~\cite{sep}. This constitutes an exceptional situation, which we will not address in detail.

The general expression for the local cargo density, Eq.~(\ref{eq:mf_rho}), in the steady state reads
\begin{equation}
(2\rho - 1)\partial_{x}\rho - ( \Omega_{de} + \Omega_{ev} )\rho + \Omega_{de} = 0 
\label{eq:totaldensity}
\end{equation}
where, $\Omega_{de}$ and $\Omega_{ev}$ are the total effective bulk deposition and evaporation rates respectively, which read
\begin{equation}
\Omega_{de} = \frac{D_0}{C_{o}},\;\;\;\;\;\;\;\Omega_{ev} = \frac{E_{+} + E_{-}}{2C_{o}} + \frac{E_{+} - E_{-}}{2}
\end{equation} 
where, $C_{0} = \frac{1- K_{+}}{1 + K_{+}}$. Eq.~(\ref{eq:totaldensity}) shows that the total density evolves as an effective TASEP-LK  model where the effective deposition and evaporation rates are modulated by the interconversion rate, $K_{+}$.
We can identify accordingly the effective  binding rate constant, $K_{\mbox{lk}} =\Omega_{de}/\Omega_{ev}$, which implies that in the absence of boundaries the cargo density relaxes to the Langmuir isotherm density, $\rho_{\mbox{lk}} = \frac{K_{\mbox{lk}}}{1 + K_{\mbox{lk}}}$.  In the absence of interconversion, and for one moving species ({\sl i.e.}  $K_{+} = E_{-} = D_{-} = 0$), $K_{\mbox{lk}}$ and $\rho_{\mbox{lk}}$  reduce to those of single species TASEP, subject to Langmuir kinetics~\cite{freylet}.

The particular case where $\Omega_{de} = \Omega_{ev}$,  $K_{\mbox{lk}}=1$, leads  to  symmetric behavior of the two motor species, where $
\frac{D_0}{C_{0}} = \frac{E_{+}+E_{-}}{2C_{0}} + \frac{1}{2}(E_{+} - E_{-}) = A$, and the overall cargo density  follows the simple expression
\begin{equation}
(2\rho - 1)\left[\frac{d\rho}{dx} - A\right] = 0. 
\label{eq:profile_symm}
\end{equation}

\subsection{Boundary Conditions}

The presence of two species moving in opposite directions alters the dynamics at the filament ends, whose behavior become relevant in the dynamical properties of ensembles of cargoes when the incoming and outgoing fluxes determine the particle distribution along the filament.  In the  thermodynamic limit  the  entry and exit rates of the vacancies at site $i = 0$, read respectively
\begin{eqnarray}
R_{en} &=& \rho_{+}(0)\left[1 - \rho_{+}(0) - \rho_{-}(0)\right] + \beta_{-}\rho_{-}(0) \\
R_{ex} &=& \alpha_{+}\left[1 - \rho_{+}(0) - \rho_{-}(0)\right] + \rho_{-}(0)\left[1 - \rho_{+}(0) - \rho_{-}(0)\right] \nonumber
\end{eqnarray}
At steady state $R_{en}=R_{ex}$. Using Eq.~(\ref{eq:ratio}), the expression for the density  of (+)-movers at the left filament end is given by,
\begin{equation}
\rho_{+}(0) = \frac{M \mp \left[M^{2} -4\alpha_{+}W_{+}\left(1 - K_{+}\right)\right]^{\frac{1}{2}}}{2W_{+}\left(1 -K_{+}\right)}
\label{eq:boundary1}
\end{equation}
where $M = \alpha_{+}W_{+} + \beta_{-}K_{+} + 1 - K_{+}$,   which can also be derived in steady state from Eq.(~\ref{eq:average_1}). Similarly,  at the other filament end, $x = 1$,
\begin{equation}
\rho_{+}(1) = \frac{P + \left[P^{2} +4\alpha_{-}W_{+}\left(1 -K_{+}\right)\right]^{\frac{1}{2}}}{2W_{+}\left(1 -K_{+}\right)}
\label{eq:boundary2}
\end{equation}
where $P = -\alpha_{-}W_{+} - \beta_{+} + 1 - K_{+}$.
The boundary densities do not depend   on  the attachment/detachment  rates and coincide with the predictions for a bidirectional TASEP  model decoupled from the environment, with $D_\pm = E_\pm =0$~\cite{madan}. Unlike  unidirectional TASEP-LK~\cite {freylet}, stochastic switching leads to  boundary densities  which do not depend linearly on the  entry and exit particle  rates. 

\subsection{Monte Carlo (MC) simulations}

In order to check the validity and limitations of the mean field approach introduced previously, we have carried out  Monte Carlo (MC)  simulations where we have implemented the dynamic processes described in Section~\ref{sec:model}. A particle on the lattice  and one of the different processes are selected at random for a MC move. A move for a particular process( i.e, translation or absorption-desorption) is accepted proportional to its rate. To achieve numerically the proper scaling between the rates associated with the different dynamical processes and to reach time scales in which cargoes can displace significantly along the filament, we take advantage of the fact that  the interconversion dynamics (or stochastic switching) between (+)- and (-)-motors is the fastest process. Hence, we enforce the time scale separation performing a large number of interconversion trial moves between each of the other two trial processes. We have checked that above  $20$ interconversion attempts pe

We start from a random distribution of particles and let the system evolve  before averaging. After an initial transient of around  $\geq 500 \frac{N}{r}$ swaps, where $r$ stands for the rate of the slowest process (either deposition/evaporation or end filament particle fluxes), the system  reaches its steady state. We then gather statistics of the relevant quantities  averaging typically  over $10^4$ time swaps and collect information with a period  $\geq 10 \frac{N}{r}$.

\section{Density and current profiles}
\label{sec:boundary}

In order to  establish the general non-equilibrium phase diagrams which characterize the collective motion of the two species of cargoes, we need to analyze the allowed  density profiles in the regime where the dynamics along the filaments compete with the  boundary  fluxes. We will discuss separately the symmetric situation where the adsorption/desorption rates are equal, $K_{\mbox{lk}}=1$, from the general case $K_{\mbox{lk}}\neq 1$. The former, simpler case will help us to  put forward the basic competing scenarios and to clarify the central role played by the incoming fluxes at the filament ends.

\subsection{ $K_{\mbox{lk}} = 1$ }
We focus on the symmetric case to set the general structure of the collective motion of different species. According to Eq.(~\ref{eq:profile_symm}),  the steady profiles read 
\begin{equation}
\rho_{+} = \frac{1}{2 W_+} \;\;\; , \;\;\; J_{+}  =  \rho_{+}(1 - \rho_{+} - \rho_{-}) =  \frac{1}{4 W_+} 
\end{equation}
where use has been made of Eq.~(\ref{eq:ratio}). The  fluxes suggests that for the homogeneous profile, $\rho=1/2$,  the total current  is maximized and thus corresponds to a maximal current phase, just like for TASEP system. We will refer subsequently to the homogeneous profile as  $\rho_{+}^{MC}$. 

The inhomogeneous density profiles of $+$ and $-$ moving cargoes  vary linearly and cannot satisfy the two boundary conditions at the same time. When the net  particle flux is positive, $K_{+} < 1$, the overall density and individual   density for both $(+)$ and $(-)$  species increases linearly. We will distinguish between profiles which satisfy the density constraint at the left (right) filament end, i.e.  low (high) density, $\rho^{LD}_+$ ($\rho^{HD}_+$). The corresponding phases will be named  accordingly as the low density (LD) and high density (HD) phases.
When the overall flux is from right to left, i.e when  $K_{+} > 1$, the solution which satisfies the density constraint on the right (left) boundary will correspond to the LD (HD) phase; the overall density profile grows from right to left. 

Without loss of generality, we will restrict ourselves to $K_{+} < 1$, because the behavior for $K_+>1$ can be derived    under the transformation $K_{+} \leftrightarrow 1/K_{+}$, $\alpha_{+}\leftrightarrow \alpha_{-}$, $\beta_{+}\leftrightarrow \beta_-$, $E_+\leftrightarrow E_{-}$, $D_{+}\leftrightarrow D_{-}$ if we take into account that the overall flux reverses direction.
Thus for $K_{+} < 1$
  \begin{eqnarray}
\rho_{+}^{LD}(x) &=& \rho_{+}(0) + Bx\\
\rho_{+}^{HD}(x) &=& \rho_{+}(1) - B(1-x)
\end{eqnarray}
where $B = D_0/(1 -K_+)$. Here, $\rho_{+}(0)$ and $\rho_{+}(1)$ are the density of the $(+)$ species at the left and the right boundaries, as prescribed by Eq.~(\ref{eq:boundary1}) and Eq.~(\ref{eq:boundary2}) respectively. Although the rates of incoming and outgoing cargoes at the filament ends are natural control parameters, it is easier to classify the  allowed coexistence regimes in terms of the  densities at the filament ends.

Let us consider a low density phase propagating from the left end of the filament and a  high density phase propagating in the opposite direction while in the filament center  a maximal current phase can develop.  If the densities at the two filament ends satisfy $\rho_{+}(0) \leq \frac{1}{2 W_+}$ and $ \frac{1}{2 W_+} \leq \rho_{+}(1) $, we can identify the two position where the corresponding currents equal the maximum current. Such positions, $x_c^{(LD/MC)}$ and $x_c^{(HD/MC)}$, respectively, determine the  stability  boundary of the corresponding phase. 
If the maximal current phase is involved, current  continuity implies also a continuous change in particle density, similar to the situation in single species TASEP. However unlike TASEP, the LD and HD profiles  are not spatially homogeneous. Using the current continuity condition, the  crossover positions can be expressed in terms of the  boundary densities  as,


\begin{eqnarray}
x_{c}^{(LD/MC)} &=& \frac{1}{B}\left[ \frac{1}{2W_+} -\rho_{+}(0)\right]\nonumber\\
 x_{c}^{(HD/MC)} &=&x_{c}^{(LD/MC)} +1+\frac{\rho_{+}(0)-\rho_{+}(1)}{B}
\label{eq:twoshock}
\end{eqnarray}
If both crossover positions satisfy $0\leq x_{c}^{(LD/MC)} \leq x_{c}^{(HD/MC)} \leq 1$, then the three regimes  coexist in the filament, and the density profile is continuous and piecewise linear for both species. We can then write down explicitly 
\begin{equation}
\rho_{+}(x)=\left\{\begin{array}[c]{ll}  \rho_{+}(0) + Bx  &, 0 \leq x \leq x_c^{(LD/MC)} \\
 & \\
\frac{1}{2W_+} & ,  x_c^{(LD/MC)}\leq x \leq x_c^{(HD/MC)} \\
 & \\
\rho_{+}(1) - B(1-x) & , x_c^{(HD/MC)} \leq x \leq 1
\end{array}
\right.
\end{equation}
If $x_c^{(LD/MC)}\leq 0$, the low density phase is expelled from the filament. The inability of the maximal current phase to fulfill  the boundary condition  at the right filament end leads to a boundary layer. In the opposite case where $x_c^{(HD/MC)}\geq 1$ the two coexisting phases are the low density and maximal current phases and a boundary layer develops at $x=1$.
Finally, if both crossover positions lie outside the filament length the maximal current phase occupies all the filament and boundary layers develop at both filament ends.
In the limiting case where the two crossover positions coincide, $x_c^{(LD/MC)}=x_c^{(HD/MC)}$, the MC phase vanishes and particle densities move continuously from the low to the high density inhomogeneous phases and we have a two-phase coexistence regime. However for $x_c^{(LD/MC)} > x_c^{(HD/MC)}$, the intervening MC profile is expelled  and ensuring current continuity for the LD and HD solution leads to  a density discontinuity; hence  the system develops a shock in the bulk, rather than at the filament ends. The shock position, from  $J_{+}(x_{s})^{LD} = J_{+}(x_{s})^{HD}$, reads
\begin{equation}
x_s = \frac{1}{2B}\left[\frac{1}{W_+} - \rho_{+}(0) - \rho_{+}(1)\right] + \frac{1}{2}
\end{equation}
with a density jump
\begin{equation}
H_s^{+} = \rho_{+}^{HD}(x_{s}) - \rho_{+}^{LD}(x_s) =  \rho_{+}(1) - \rho_{+}(0) - B
\end{equation} 
provided  $0\leq x_s\leq 1$.

\subsection {$K_{\mbox{lk}} > 1$}

\begin{figure}
\begin{center}
\includegraphics[width=3 in,angle=-90]{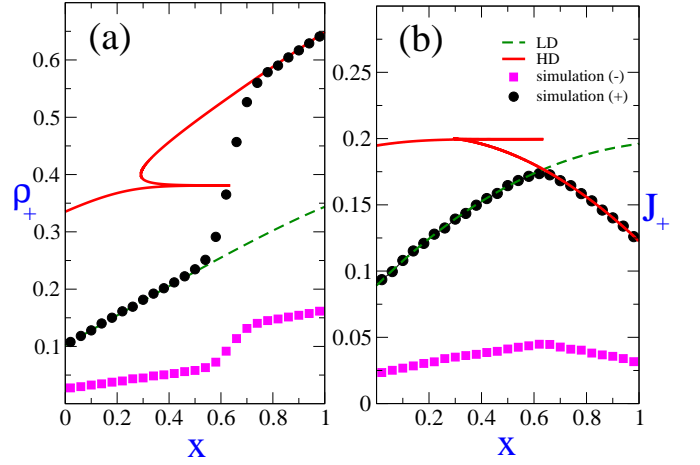}
\end{center}
\caption{(Color online)  Steady state spatial distribution of cargo density and current in the (LD-HD) phase coexistence regime, which exhibits a shock. Here $\alpha_{+} = 0.1$, $\beta_{+} = 0.2$, $\alpha_{-} = 0.2$, $\beta_{-} = 0.8$ and $D_{+} = 0.1$, $E_{+} = 0.2$, $D_{-} = 0.1$, $E_{-} = 0.3 $ and $K_{+} = 0.25$. The dots represent the profile obtained by numerical simulations for (+)-moving cargoes  while the squares are the profiles obtained by numerical simulation for $(-)$-moving cargoes  for a  system size, $N = 1000$. The lines represent the mean field  solutions of the corresponding densities and current for $(+)$-moving cargoes which satisfy the appropriate boundary condition.}
\label{figshock}
\end{figure} 
Let us analyze the general situation when the symmetry between effective deposition and evaporation rates is broken.  There exist two complementary strategies to  determine the steady state density and current profiles. One can  rewrite   Eq.~(\ref{eq:totaldensity}) in terms of $\rho_+$, 

\begin{equation}
\left[2 W_{+}\rho_{+} - 1\right]\frac{d\rho_{+}}{dx} - G\rho_{+} + B = 0 
\label{eq:rhoplus}
\end{equation} 
where $G =D_0/(C_0 \rho_{\mbox{lk}})$. Integrating Eq.~(\ref{eq:rhoplus}) and using the appropriate boundary condition  at $x = 0$  leads to
\begin {equation}
\frac{2W_{+}\left[\rho_{+}(x)- \rho_{+}(0)\right] }{G}+ \frac{2W_{+} B - G}{G^2}\log\left[\frac{G\rho_{+}(x) - B}{G\rho_{+}(0) - B}\right] = x
\end{equation}
and similarly the  density profile consistent with the right boundary condition at $x = 1$ reads,
\begin {eqnarray}
&&\frac{2W_{+}}{G}\left[\rho_{+}(x)-\rho_{+}(1)\right] + \frac{2W_{+}B - G}{G^2}\log\left[\frac{G\rho_{+}(x) - B}{G\rho_{+}(1) - B}\right]\nonumber\\
&&= x-1
\end{eqnarray}

As usual, the boundary densities $\rho_{+}(0)$ and $\rho_{+}(1)$ are  obtained from Eq.~(\ref {eq:boundary1}) and Eq.~(\ref{eq:boundary2}) once the  entry and exit rates for the two species are specified.  Fig.~\ref{figshock} compares the   MF density and current profiles and the corresponding MC simulation for a particular example, which shows density shocks in the bulk.

It is however worthwhile to  analyze the allowed steady states  in terms of the Lambert, $W(x)$, function~\cite{freypre}. This approach, which is discussed in Appendix A,  is very useful in subsequently plotting and analyzing the resultant phases for our model. \\
Depending on the boundary conditions, there exist three types of phases expressed in terms of the total density, $\rho (x)=\rho_{+}(x) + \rho_{-}(x)$:  a LD phase, $\rho_L(x)$, with the density in the bulk is dictated by the left boundary condition;  a HD phase, $\rho_R(x)$, and  an analogous to the MC phase.  As for the symmetric case, these phases may coexist, but the density profiles are no longer linear. Rather, the steady profiles  approach monotonically the Langmuir isotherm density, $\rho_{\mbox{lk}}$. $\rho_{L}(x)$ never increases beyond $1/2$, where the total current attains its maximal value. Similarly  the solution, $\rho_{R}(x)$, always greater than $1/2$, approaches the Langmuir isotherm density, $\rho_{\mbox{lk}}$, becoming independent of the boundary conditions from above (below) depending  on the  density of right moving motors,  $\rho_{R}(1)>\rho_{\mbox{lk}}(\rho_{R}(1)<\rho_{\mbox{lk}})$~\cite{example}. Due to these properties, there are different scenarios depending on the  imposed overall densities at the filament ends:\\ 

If $\rho_{L}(0)< 1/2, \rho_{R}(1)> 1/2$, both solutions match the respective boundary conditions. HD and  LD phases coexist  and $\rho(x)$  is obtained locating the position $x_w$ where the fluxes cross each other and  the   domain wall develops.  If $x_w > 1 (x_w <0)$, no coexistence is allowed and the  filament density is determined only by  the left (right) filament end. \\

When $\rho_{L}(0)< 1/2, \rho_{R}(1)< 1/2$, the density profile  matching the right boundary condition is unstable and  the  density approaches the extremal solution $W_{0}(-Y_{*}(x))$, described in Appendix A, approaching 1/2 at the right filament end. This phase, referred to as  (M) phase~\cite{freypre},  is analogous to the maximal current (MC) phase  described previously; the filament density is independent of the boundary value and the current is maximal. Unlike TASEP, the M phase has a spatially varying density profile , with the current approaching the extremal value only at the filament end.  \\
If  $\rho_{L}(0) > 1/2$, the bulk profile is already in the high density phase. However, for $\rho_{R}(1) < 1/2$, the right solution is given  as before by its extremal value and the  entire bulk phase density ends in   the M phase profile because  matching the right boundary condition is unstable. As a result, the profile at the right boundary approaches the extremal solution $W_{0}(-Y_{*}(x))$.\\

\section{Phase diagrams}          
\label{sec:phase}

The analysis of the previous section provides a systematic procedure to determine the allowed phases  and    derive the general non-equilibrium phase diagram for  collective transport of competing cargo assemblies. Since the fluxes at the filament boundaries control the   fluxes along the filament, the corresponding cargo rates are the natural parameters to describe the phase diagram. The absence of  particle/hole symmetry  leads to a non-linear relationship between particle fluxes and densities,  introducing significant differences in the phase diagrams with respect to those of single species TASEP. \\
We will focus  on phase diagram cuts in  the  $(\alpha_{+},\beta_{+})$ plane. We will discuss the main  features of these phase diagrams and compare them with  related lattice models.  Following the analysis  of the previous section, we will first discuss the symmetric limit  when the effective adsorption and desorption rates are equal. The insight gained will help us  in addressing the general scenario.

\subsection{$K_{\mbox{lk}} = 1$}
\subsubsection{\bf The phase boundaries}
\label{sec:subsect_klk}

In the previous section we have identified the  three relevant phases (LD,MC,HD), which can either cover the whole filament or coexist with each other. Hence, the  non-equilibrium phase diagram can be built once we determine the limits of coexistence of these phases. To this end, we must  determine the position of  $x_c^{(LD/MC)}$, $x_c^{(HD/MC)}$ and the position of shock, $x_s$ and analyze when they are expelled from the filament. Hence we discuss  the  coexistence onset between the different phases in terms of the  boundary densities. The corresponding expressions showing the explicit dependence on the boundary rates $\alpha_{+}$ and $\beta_{+}$ are shown in Appendix B.

{\sl Phase coexistence line between (LD-HD) and (LD-MC-HD):}  MC phase is expelled when  $ x_c^{(LD/MC)} > x_c^{(HD/MC)} $. Therefore, LD and HD phases will start to coexist with an MC phase when  the  density shock will vanish, {\sl i.e.} $x_c^{(LD/MC)} = x_c^{(HD/MC)}$  along the filament. In terms of the boundary densities, from   Eq.~(\ref{eq:twoshock}) it follows  
\begin{equation}
\rho_{+}(1) - \rho_{+}(0) = B,
\label{eq:pb1}
\end{equation} 
from which the phase boundary in $(\alpha_{+}, \beta_{+})$ plane  is obtained by using Eqs.~(\ref{eq:boundary1}) and (\ref{eq:boundary2}), which relate the boundary densities $\rho_{+}(1)$ and  $\rho_{+}(0)$ with $\alpha_{+}$ and $\beta_{+}$. The obtained expression has two possible solutions depending on the parameters. Determining the physically  plausible solution cannot always be resolved; we have then used Monte Carlo simulations to identify the appropriate boundary line. 

{\sl Phase coexistence line between  (LD)  and  ( LD-HD ): }
When the shock position, $x_s$ , is  close to the right filament end, then the profile in the filament is essentially  LD solution with an incipient  HD profile. Accordingly, the  LD-HD  density shock starts at $x_s = 1$. In terms of the boundary density, this condition reads
\begin{equation}
\rho_{+}(1) + \rho_{+}(0) = \frac{1}{W_+} - B
\label{eq:pb2}
\end{equation}

{\sl Phase coexistence line between   (HD)  and ( LD-HD ) :}
Analogously to the preceding case,  phase coexistence will start with a density shock at  $x_s = 0$, 
\begin{equation}
\rho_{+}(1) + \rho_{+}(0) = \frac{1}{W_+} + B
\label{eq:pb3}
\end{equation}

{\sl Phase coexistence line between  (LD) and  ( LD-MC ) and  between (LD-MC) and (MC):}
When $x_c^{(LD/MC)}$ is close to 1(0)  the bulk profile corresponds to an incipient   LD (MC) at the left  (right) boundary and joining up with an emerging MC (LD) profile at the  right (left)  boundary.  Therefore,  the position of the coexistence line is determined by the limiting expression  $x_c^{(LD/MC)} = 1(0)$, so that the equations of phase boundaries are, 
\begin{eqnarray}
\rho_{+}(0) &=& \frac{1}{2W_+} - B\label{eq:pb4}
\\
\rho_{+}(0) &=& \frac{1}{2W_+}
\label{eq:pb5}
\end{eqnarray}
for  (LD)-( LD-MC ) and (LD-MC)-(MC), respectively. As shown in Appendix B,  the  coexistence curve is a straight line  with fixed $\beta_{+}$.

{\sl Phase coexistence curve between   (HD) - (HD-MC) and between  (HD-MC) - (MC):} The coexistence starts when the MC phase is allowed at the right (left)  filament end; i.e.   $x_c^{(HD/MC)} = 0$ ($x_c^{(HD/MC)} = 1$). Hence,
\begin{eqnarray}
\rho_{+}(1) &=& \frac{1}{2W_+}
\label{eq:pb6}
\\
\rho_{+}(1) &=& \frac{1}{2W_+} + B
\label{eq:pb7}
\end{eqnarray}
for  (HD)-(HD-MC) and (HD-MC)-(MC) coexistence onset, respectively.

\begin{figure}
\begin{center}
\includegraphics[width=2.5in,angle=-90]{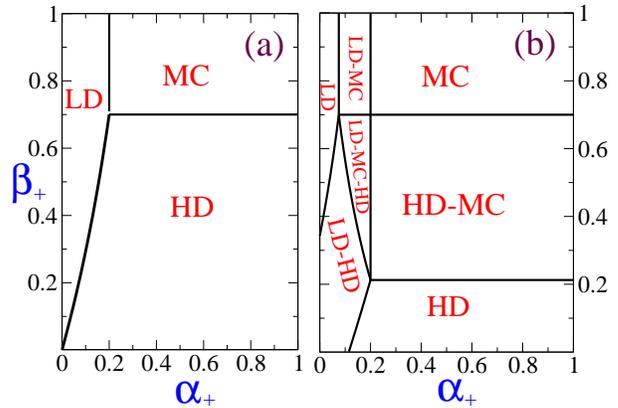}
\end{center}
\caption{(Color online)  Phase space cut along the $\alpha_+$-$\beta_+$ plane for equal effective deposition and evaporation cargo  rates ( $K_{\mbox{lk}} = 1$). (a) Without LK: When deposition and evaporation processes are not present then two or three phase-coexistence is absent and the phase diagram reduces to the case of two species TASEP with stochastic switching. Here $\beta_{-} = 0.1$ , $\alpha_{-} = 0.3$, $K_{+} = 0.5$. (b)  LK is introduced, with $D_{0} = 0.1$ and the rest of  parameters as before. There are clear topological changes in the phase diagram, with new regions of phase coexistence opening up.}
\label{madan}
\end{figure}
\begin{figure}
\begin{center}
\includegraphics[width=2.5in,angle=-90]{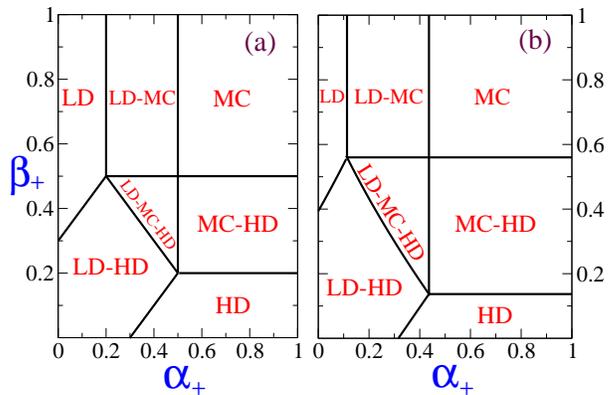}
\end{center}
\caption{(Color online)  Phase space cut along the $\alpha_+$-$\beta_+$ plane for equal effective deposition and evaporation cargo  rates ( $K_{\mbox{lk}} = 1$).(a) Without interconversion process ($K_{+}= 0$): Here $\beta_{-} = 0$ , $\alpha_{-} = 0$, $D_{+} = 0.3$, $E_{+} = 0.3$, $D_{-} = 0$, $E_{-} = 0$. In this limit the phase diagram reduces to the case of single species TASEP-LK and the phase diagram corresponds to the ones reported in Ref.~\cite{freypre}. (b) $K_{+}= 0.1$, $\alpha_{-} = 0.1$, $\beta_{-} = 0.3$ and $D_{0} = 0.3$ and holding $K_{\mbox{lk}} = 1$ . The  phase diagram structure changes significantly from the previous case, with the LD region shrinking, accompanied by change in shape of (LD-HD) coexistence region. Also in this case the phase coexistence line between (LD-MC-HD) with (LD-HD) is not a straight line leading to the possibility of a {\it re-entrant behaviour} (see text).}    
\label{frey}
\end{figure}

\subsubsection{\bf Nature of the phase diagram}
Using the  expressions for the coexistence curves, we can  analyze the  phase diagrams which emerge as a result of the transport of two competing species.\\
The presence of two species moving in opposite directions allow for overall current inversion as a function of the interconversion rate, a feature  present in the absence of LK~\cite{madan}. In Fig.~\ref{madan} we compare the effect of LK  in the collective cargo transport  with respect to collective transport in the absence of  LK. One can appreciate that interconversion prevents particle-hole symmetry, as observed above; moreover, LK is necessary to allow phase coexistence. Further, in the absence of LK the density and current profiles  are homogeneous along the filament, unlike the ones corresponding to Fig.~\ref{figshock}.\\
In Fig.~\ref{frey} we show the phase diagram  in terms of the  incoming and outgoing rates of $(+)$ species at the filament left and right ends, for a particular set of adsorption and desorption rates.  As anticipated, by varying the particle fluxes we can observe either  density profiles corresponding to a single phase or coexistence of two or three different phases. Fig.~\ref{frey}a shows collective transport in the absence of interconversion, when the model becomes equivalent to unidirectional TASEP-LK if  $\alpha_-=\beta_-=0$.  Comparing it with Fig.~\ref{frey}b, one can appreciate that the breakdown of particle-hole symmetry induced by stochastic switching leads to a lack of symmetry in the phase diagram along the diagonal axis, $\alpha_{+} = \beta_{+}$.  This peculiarity leads to qualitatively new scenarios when motor species switch stochastically. \\
 For example,  the phase boundary between (LD-HD)-(LD-MC-HD), shown in Fig.~\ref{frey}b and Fig.~\ref{madan}b are no longer straight lines, as is clear from Eq.(\ref{eq:betaplus}). Hence, phase {\it re-entrance} is allowed; when moving along a linear path  decreasing $\alpha_+$ and increasing accordingly $\beta_+$, it is possible to pass from coexistence between a LD and HD profiles with a  shock to a three phase coexistence (LD-MC-HD) and finally re-emerge into LD-HD phase coexistence, exhibiting a re-entrant density shock~\cite{errata}. As shown in Fig.~\ref{frey}b and Fig.~\ref{madan}b,  topologically  the phase boundary curve between (LD-HD)-(LD-MC-HD) region always ends at the vertex of another phase boundary curve.  

\subsection{$K_{\mbox{lk}} > 1$}
We will follow the approach carried out in the previous subsection and analyze first  the coexistence boundary curves and   analyze subsequently the main features of the  corresponding phase diagrams.

\begin{figure}
\begin{center}
\includegraphics[width=2.7in,angle=-90]{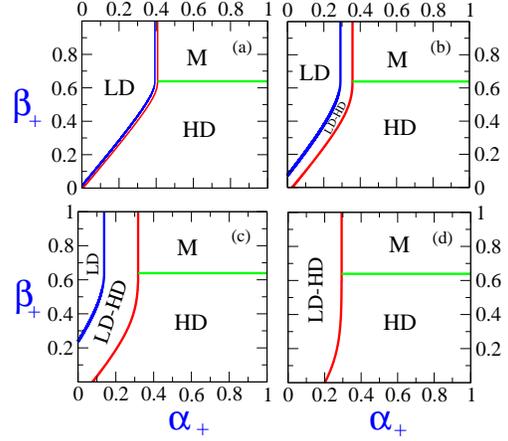}
\end{center}
\caption{(Color online)  Phase space cut along the $\alpha_+$-$\beta_+$ plane varying the  total deposition cargo  rates, $D_{0} \equiv$ $(D_{+} + D_{-})$, for fixed value of $K_{\mbox{lk}} = 2$ and $K_{+}=0.2$. (a) $D_{0}$ = 0.02, (b) $D_{0} = 0.1$, (c) $D_{0} = 0.3$, (d) $D_{0} = 0.9$. $E_{+} = E_{-}$, $\alpha_{-} = 0.2$ and $\beta_{-}= 0.8$. For very low values of the total deposition rate as in (a), the topology of the phase diagram is similar to the one obtained in the absence of LK, where phase-coexistence is  less favoured.}
\label{Fig5}
\end{figure}
\begin{figure}
\begin{center}
\includegraphics[width=2.7in,angle=-90]{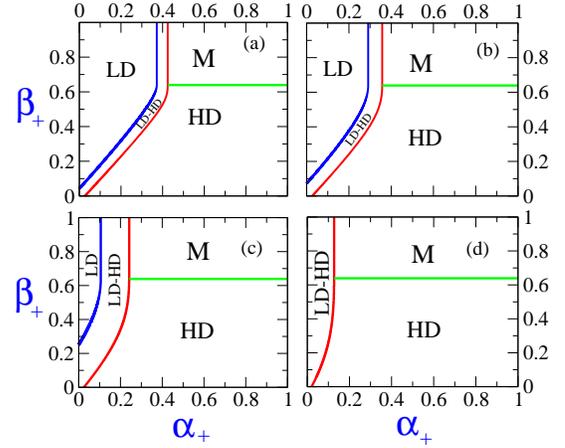}
\end{center}
\caption{(Color online)  Phase space cut along the $\alpha_+$-$\beta_+$ plane varying $K_{\mbox{lk}}$ for fixed value of $K_{+}=0.2$, $E_{+}$ = $E_{-} = 0.05$ and  $\alpha_{-} = 0.2$ and $\beta_{-}= 0.8$. (a) $K_{\mbox{lk}} = 1.2$, (b)$K_{\mbox{lk}} = 2$, (c) $K_{\mbox{lk}} = 6$ and (d) $K_{\mbox{lk}} = 18$. Increasing  $K_{\mbox{lk}}$ decreases  the low density region  until it is  eventually expelled; larger $K_{\mbox{lk}}$ implies higher bulk cargo density.}
\label{Fig6}
\end{figure}
\begin{figure}
\begin{center}
\includegraphics[width=2.7in,angle=-90]{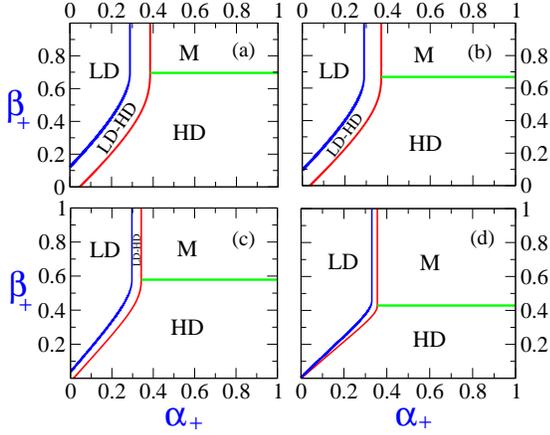}
\end{center}
\caption{(Color online)  Phase space cut along the $\alpha_+$-$\beta_+$ plane varying the interconversion rate constant, $K_{+}$, for fixed values of $K_{\mbox{lk}}= 2$, $E_{+}$ = $E_{-} = 0.05$ and $\alpha_{-} = 0.2$ and $\beta_{-}= 0.8$. (a) $K_{+} = 0.01$, (b)$K_{+} = 0.1$, (c) $K_{+} = 0.4$ and (d) $K_{+} = 0.8$.  The maximal current phase  region is favored on increasing  $K_{+}$. The shape of the LD-HD coexistence region differs significantly  from the shapes observed in the absence of interconversion~ \cite{freypre}.}
\label{Fig7}
\end{figure}
\begin{figure}
\begin{center}
\includegraphics[width=2.7in,angle=-90]{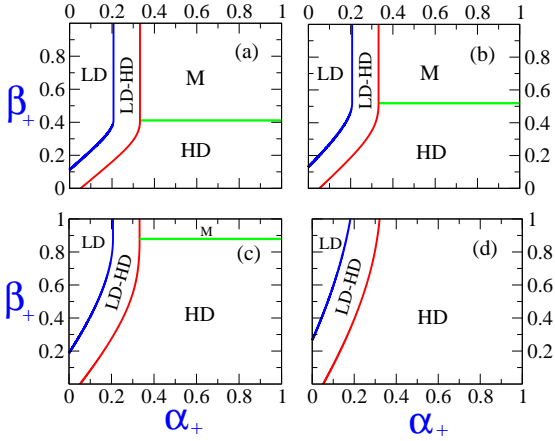}
\end{center}
\caption{(Color online) Phase space cut along the $\alpha_+$-$\beta_+$ plane varying the entry cargo rate, $\alpha_{-}$, of $(-)$-moving  species for fixed values of $K_{\mbox{lk}} = 2$, $K_{+}=0.2$, $\beta_{-}= 0.8$ and $D_{0} = 0.2$. (a) $\alpha_{-} = 0.01$, (b)$\alpha_{-} = 0.1$, (c) $\alpha_{-} = 0.4$ and (d) $\alpha_{-} = 0.8$. For high $\alpha_{-}$, system never attains current saturation for higher value of $\beta_{+}$ and in fact the M phase is absent, a feature missing in  single species TASEP and TASEP with LK~\cite{freypre}.}
\label{Fig8}
\end{figure}

\begin{figure}
\begin{center}
\includegraphics[width=2.4in,angle=-90]{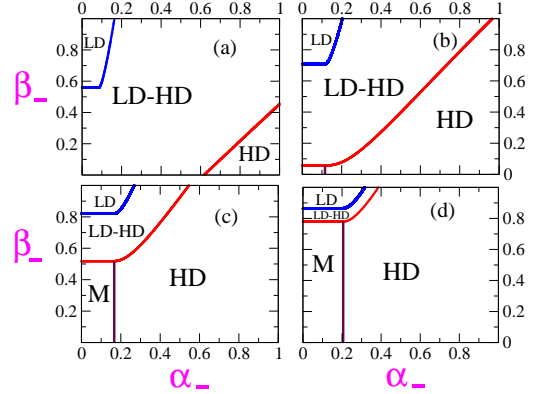}
\end{center}
\caption{(Color online)  Phase space cut along the $\alpha_-$-$\beta_-$ plane varying the interconversion rate constant , $K_{+}$,   for fixed values of $K_{\mbox{lk}} = 1.1429$,  $\alpha_{+} = 0.25$, $\beta_{+}= 0.5$ and $D_{0} = 0.2$. (a) $K_{+} = 0.2$, (b)$K_{+} = 0.3$, (c) $K_{+} = 0.5$ and (d) $K_{+} = 0.7$. For low  $K_{+}$ the system never attains current saturation and the M phase is expelled from the phase diagram}
\label{Fig9}
\end{figure}

\subsubsection{ \bf Phase boundaries}

Three phase coexistence is not allowed now  due to the different nature of the maximal current phase.  Therefore, the relevant curves reduce to three possibilities.

{\sl Phase coexistence line between  (LD) and (LD-HD):} The incipient coexistence  develops when the fluxes of the two phases match  at the right filament end, leading to
\begin{equation}
\rho_{L}(1) + \rho_{R}(1) = 1
\end{equation} 
Relating  the right boundary density with the particle input and output rates using Eq.~(\ref{eq:boundary2}), one arrives at
 \begin{equation}
\beta_{+} = \rho_L(1)\left[1-K_++\frac{\alpha_-W_+}{1-\rho_L(1)}\right]
\label{eq:pbg1}
 \end{equation}
 in terms of  $\rho_{L}(1)$. Using Eqs.~(\ref{eq:boundary1}) and (\ref{eq:sigma}) one can derive numerically $\rho_L(1)$ as a function of the input fluxes $\alpha_{+}- \beta_{+}$. 
 
 After obtaining $\rho_{L}(0)$ using Eq.~(\ref{eq:boundary1}), $Y_{L}(x = 1)$ is evaluated. Using this, Eq.~(\ref{eq:sigma}) is solved numerically to obtain $\sigma_{L}(1)$ and hence $\rho_{L}(1)$. Thus the entire phase boundary line in the $\alpha_{+}- \beta_{+}$ plane can be constructed using these set of relations.  

{\sl Phase coexistence line between   (HD) and (LD-HD):}  
Phase coexistence starts when the density shock appears at the left filament end, $x_s=0$, leading to
\begin{equation}
\rho_{L}(0) + \rho_{R}(0) = 1
\end{equation} 
 Using the relation of the left boundary density with the particle input and output rates through, Eq.~(\ref{eq:boundary1}), one obtains the relation,
\begin{equation}
\alpha_{+} =  \frac{T^{2} - 2T(1 -K_{+}  + \beta_{-}K_{+} ) }{2TW_{+} - 4( 1 - K_{+}^{2})}
\label{eq:pbg2}
\end{equation}
where $T =  2(1 - K_{+})\rho_{L}(0)$ and we follow an analogous  procedure to the previous case to relate $\rho_{R}(0)$  to the incoming and outgoing cargo rates, $\alpha_{+}$ and $\beta_{+}$.
 
{\sl Onset of the  M phase:}
The maximal current phase sets in when the maximum LK density, $\rho_{R}(1) = 1/2$, develops at the right filament end. The corresponding value for the entry rate reads
\begin{equation}
\beta_{+} = \alpha_-W_++\frac{1+K_+}{2}.
\label{eq:pbg3}
\end{equation} 

\subsubsection{\bf Nature of the phase diagram}
The topology of the phase diagram and even the nature of the phases are  quite distinct from the symmetric case because three phase coexistence is not allowed. Moreover, the phase coexistence lines separating LD and HD have a different functional form and are not parallel any more. The nature of the M phase, although insensitive to the boundary rates, is no longer homogeneous and it is not characterized by a constant  flux density. \\
Figs.~\ref{Fig5} and Fig.~\ref{Fig6} illustrate how LD phase becomes less favored and  it is eventually expelled on increasing the overall deposition, $D_{0}$, or $K_{\mbox{lk}}$, leading to a  topological change of the phase  diagram. Such changes leave the M-HD coexistence curve  virtually unaffected. Modifying instead   the interconversion, $K_{+}$, and (-)-cargo  entry, $\alpha_{-}$, rates affect the position of the M-HD phase boundary as shown in Fig.~\ref{Fig7} and Fig.~\ref{Fig8}. While always present in one species TASEP-LK~\cite{freypre}, the maximal M phase is destabilized on increasing   $\alpha_{-}$, and  beyond a critical value it is expelled from the phase diagram, modifying its topology as shown in Fig.~\ref{Fig8}d. Interconversion enhances significantly the sensitivity of the distance  between the (LD)-(LD-HD)  and (HD)-(LD-HD)  enhancing the corresponding coexistence regions, as displayed in Figs.~\ref{Fig7}a and \ref{Fig7}b. Fig.~\ref{Fig7} shows how   distance between these two coexistence  lines varies significantly and vanishes  for very small values of $\alpha_+$ and $\beta_+$. 

\subsubsection{\bf Phase diagram in terms of incoming and outgoing rate of $( - )$ species}
The expressions derived above describing the phase boundaries can be used to describe the phase diagram in terms of different control parameters. For example, if we use the entry rates of the $(-)$ motors, we arrive at analogous  phase diagrams, as shown, for example, in Fig.~\ref{Fig9}. Although similar to the phase diagrams in terms of the entry rates of the $(+)$ species, the phase boundaries are rotated. For example in Fig.~\ref{Fig9}  the bottom right corner is HD phase, the bottom left is M phase,  the middle region is essentially a LD-HD coexistence region and the  top left corner is LD phase.  
Due to the symmetry between the two motor species, the phase diagram in  $\alpha_{+},\beta_{+}$ is identical  to the corresponding one in the  $\alpha_{-},\beta_{-}$  plane for the transformation $K_{+}\rightarrow K_{-}$ ($K_{-} =k_{-}/k_{+} $),    $E_{+}\rightarrow E_{-}$,  $D_{+}\rightarrow D_{-}$,  $E_{-}\rightarrow E_{+}$ and  $D_{-}\rightarrow D_{+}$, although with a reversed overall motor flux density. It is worth noting that the diagrams in the  $\alpha_{+},\beta_{+}$  plane for  $K_{+} < 1 $ are related by  mirror inversion and a $90^o$ rotation,  with the corresponding diagrams in the    $\alpha_{-},\beta_{-}$  plane for  $K_{+} <1 $ ( and equivalently with $\alpha_{+},\beta_{+}$ plane  for  $K_{+} > 1 $). This symmetry is appreciated by comparing the phase plane plots of  Figs.~\ref{Fig8} and   \ref{Fig9}. 

\section{Conclusions}
\label{sec:disc}
We have presented a minimal model for bidirectional cargo  transport on biofilaments to address the basic features associated to their collective dynamics. We have identified the relevant  competing rates, the central role played by motor regulation,  and  have  described the expected scenarios depending on  their relative  magnitudes.
We have shown that despite their opposed directionality, the collective transport may favor one particular global direction, in agreement  with experimental  observations. We have analyzed how current reversal emerges and its impact in cargo density accumulation at either end of a polar filament.  We have focused on the regime where interconversion is the rate limiting process, a regime consistent with experimental observation of vesicle transport~\cite{gross_06}. The rate of  interconversion may affect the phase diagram shape, as mentioned in Appendix C, and the framework described can be easily adapted to analyze such regimes. \\
This  'fast' interconversion regime, where the interconversion is the rate limiting process is simpler in nature and it has served to pinpoint the main features of collective transport of opposed polarity species. The derived non equilibrium phase diagrams rationalize the interplay between  motor regulation and motor exchange and they exhibit interesting and rich behavior and shows the relevant role played by motor regulation when several motor species interact to promote active cargo transport. In particular, we have described how shock re-entrance is allowed, and also how interconversion destabilizes and prevents certain phases for appropriate parameters. The interplay of the different  processes  determining collective cargo motion provides a large degree of flexibility in phases and structures which cargoes can develop as compared to  cargoes pulled by one molecular motor species. For example, both the maximal and low density phases can be expelled, and the regimes of LD and HD profiles  become more  common as compared to  one species TASEP. Moreover, the regulation of the two opposed species can lead to flux reversal without modifying the underlying phase diagram.

Despite the simplicity of the model, we can relate the relevant model  coefficients $K_+$, $K_{\mbox{lk}}$ and $D_{0}$ to experimentally relevant situations.  {\it In-vitro} experiments with cellular extracts of MT and molecular motors,  motor regulation (and thus effectively $K_{+}$) is achieved controlling dynein cofactors  p150/{\sl Glued} and  {\sl dynamitin} concentration, which are subunits of the  {\sl dynactin}  and regulate the relative binding affinities of dyneins  to the MT~\cite {welte}. Experimentally it is known that increasing the cofactor concentration leads cargo motion reversal~\cite {welte, borisy1}.  $K_{\mbox{lk}}$ and $D_{0}$ can be related to the motor interaction with the filament and can be controlled by modifying either motor processivity~\cite{borisy1,borisy2}, filament concentration or the attachment properties of the cargo to other filaments.

In order to  address the   competition between different mechanisms and how they affect the collective properties of cargo transport, we have analyzed how  different phases develop along a filament  as the attachment/detachment, interconversion and boundary input(output) rates vary.  In particular,  we have shown how one can move from a state of (LD-HD) phase coexistence, characterized by existence of density shocks,  to states with smooth density profile varying either the interconversion rate, $K_{+}$, or the effective desorption rate, $D^{*}$ . Therefore, the  interplay between two  different regulatory mechanisms, the detachment/attachment processes( e.g. due to the presence of an embedding actin network) and interconversion due to motor regulation, can give rise to a rich variety of  inhomogeneous collective cargo distributions. Understanding the impact of such scenarios in realistic biological systems remains an open challenge.

\acknowledgments
IP acknowledges Spanish MICINN, Project FIS2008-05386, and Generalitat de Catalunya (DURSI, SGR2009-00634), for financial support.

\appendix

\section{Analysis of steady states using Lambert function}

It is useful to  analyze the allowed steady states  introducing  the auxiliary function $\sigma (x)$~\cite{freypre},
\begin{equation}
\sigma(x) = \frac{K_{\mbox{lk}} + 1}{K_{\mbox{lk}} - 1}[2\rho(x) - 1] - 1
\label{eq:rhosig} 
\end{equation}
where $\rho (x)=\rho_{+}(x) + \rho_{-}(x)$ refers to the total density.
Expressed in terms of $\sigma (x)$, Eq.~(\ref{eq:totaldensity}) can be integrated, leading to 
\begin{equation}
|\sigma(x)|\exp[\sigma(x)] = Y(x)
\label{eq:sigma}
\end{equation}
where,
\begin{equation}
Y(x) = |\sigma(x_{0})|\exp \left[ \Omega_{ev}\frac{{(K_{\mbox{lk}} + 1)}^{2}}{K_{\mbox{lk}} -1}(x - x_{0})+\sigma(x_{0})\right]
\label{eq:Y}
\end{equation}
Here, $x_{0}$, refers to the position of the left (right )boundary and $\sigma(x_{0})$ is the corresponding value of the function at that boundary.\\

Eq.~(\ref{eq:sigma}) has a explicit solution in terms of  the Lambert, $W(x)$, function~\cite{freypre}. 
\begin{equation}
\sigma(x) = \left\{\begin{array}[c]{ll}
W(Y(x)) & , \sigma(x) > 0 \\
 & \\
 W(-Y(x)) & , \sigma(x) < 0
\end{array}
\right.
\end{equation}
The Lambert  function has two real branches, $ W_{0}(Y(x))$ and $W_{-1}(Y(x))$, in terms of which we can write

\begin{equation}
\sigma(x) = \left\{\begin{array}[c]{lll}
 W_{-1}(-Y) & : \rho \in [0,\frac{1}{2}] & , \sigma\in [\frac{-2K_{\mbox{lk}}}{K_{\mbox{lk}}-1}, -1] \\
  & & \\
 W_{0}(-Y) & :  \rho \in [\frac{1}{2},\frac{K_{\mbox{lk}}}{K_{\mbox{lk}} +1}] & , \sigma\in [-1,0] \\
  & & \\
 W_{0}(Y) & :   \rho \in [\frac{K_{\mbox{lk}}}{K_{\mbox{lk}} +1},1] & , \sigma\in [0,\frac{2}{K_{\mbox{lk}}-1}]
 \end{array}
\right.
\end{equation}
The relevant branch of the function is selected through the boundary conditions. Specifically,  if  the  density at the left boundary lies in the  range $ 0 < \rho(0) < 1/2 $, then
\begin{equation}
\sigma_{L}(x) = W_{-1}(-Y_{L}(x)) < 0
\end{equation}
while the solution which satisfies the right boundary condition at $x = 1$ fulfills
\begin{equation}
\sigma_{R}(x) = \left\{\begin{array}[c]{ll}
W_{0}(Y_{R}(x)) > 0 & , \rho_{R}(1) > \rho_{\mbox{lk}} \\
 & \\
W_{0}(-Y_{R}(x)) < 0 &, \rho_{R}(1) < \rho_{\mbox{lk}}
\end{array}
\right.
\end{equation} 
where $Y_{L}(x)$ and $Y_{R}(x)$ are the functions  $Y(x)$ introduced in  Eq.~(\ref{eq:Y}) obtained by setting $x_0 = 0$ and  $x_0 = 1$ respectively.

\section{Coexistence boundary lines}
\label{sec:appendixB}

In this appendix we provide the  explicit expressions for the coexistence boundary lines in terms of the entry and exit cargo rates   $\beta_{\pm}$ and $\alpha_{\pm}$  for symmetric TASEP-LK, where $K_{\mbox{lk}}=1$. In this way, we complement the information discussed in Section~\ref{sec:subsect_klk}, providing explicit expressions for the phase boundary lines used to construct the phase diagrams shown in Fig.\ref{frey} and Fig.\ref{madan}. In order to proceed, it is convenient to express the boundary lines in terms of the following parameters:
\begin{eqnarray}
d &=& \frac{D_0}{1-K_{+}} \nonumber\\
V &=& 4 \alpha_{-}(1-K_{+}^{2})\nonumber\\
B_{1} &=& 1- K_{+} + \beta_{-} K_{+}\nonumber\\
C &=& 1- K_{+}= \frac{D_0}{d}\nonumber\\
Q &=& 4(1-K_{+}^{2})=\frac{V}{\alpha_-}\nonumber\\
Z_{1} &=& 2(d + \rho_{0})(1-K_{+}^{2})\nonumber\\
Z_{2} &=& 2(\frac{1}{W_+}- d - \rho_{0})(1-K_{+}^{2})=-Z_1+\frac{Q}{2W_+}\nonumber\\
Z_{3} &=& 2(\frac{1}{W_+} + d - \rho_{0})(1-K_{+}^{2})\nonumber\\
Z_{4} &=& 2(\frac{1}{2W_+})(1-K_{+}^{2})\nonumber\\
Z_{5} &=& 2(d + \frac{1}{W_+})(1-K_{+}^{2})=Z_4+\frac{dQ}{2}\nonumber\\
\label{eq:ap1}
\end{eqnarray}
Here $\rho_{0} = \rho_{+}(0)$, which is given by Eq.~(\ref{eq:boundary1}). Since $\rho_{+}(0)$ depends only on entry rate of $(+)$ particle and exit rate of $(-)$ particles, $Z_{1}$, $Z_{2}$,  $Z_{3}$, $Z_{4}$ and  $Z_{5}$ do not depend on $\beta_+$.
\begin{enumerate}
\item {\sl Phase boundary between (LD-MC-HD ) and (LD-HD)}: We solve Eq.~(\ref{eq:pb1}) using the expression for the boundary densities in terms of the  incoming and outgoing rates, Eqs.~(\ref{eq:boundary1}) and  (\ref{eq:boundary2}). The relevant rates are  $\beta_+$ and $\alpha_{+}$. 
The phase boundary  line in $(\alpha_{+}, \beta_{+})$ plane is given by,             
\begin{equation}
\beta_{+} = (1-K_{+} -\alpha_{-}W_{+}) + \frac{V - Z_{1}^{2}}{2Z_{1}},
\label{eq:betaplus}
\end{equation}
Note that the right hand side of Eq.~(\ref{eq:betaplus}) is an explicit function of $\alpha_{+}$. Unlike the symmetric case,  $K_{\mbox{lk}} = 1$, the entry and exit rate, $\beta_{+}$ and $\alpha_{+}$, are not proportional to each other.  

\item {\sl Phase boundary between (LD-HD ) and (LD)}: In this case, the emergence of coexistence is determined by Eq.~(\ref{eq:pb2}) using the explicit  expressions for the boundary densities (Eqs.~(\ref{eq:boundary1}) and   (\ref{eq:boundary2})). 
\begin{equation}
\beta_{+} = (1-K_{+} -\alpha_{-}W_{+}) + \frac{V - Z_{2}^{2}}{2Z_{2}},
\label{eq:betaplus_1}
\end{equation}

\item {\sl Phase boundary between (LD-HD) and (HD)}: The coexistence curve is in this case encoded in  Eq.~(\ref{eq:pb3}) and on the expression of boundary densities as before.
The explicit expression is given by,
\begin{equation}
\beta_{+} = (1-K_{+} -\alpha_{-}W_{+}) + \frac{V - Z_{3}^{2}}{2Z_{2}},
\label{eq:betaplus_2}
\end{equation}   
   
\item {\sl Phase boundary between (LD) and (LD-MC)}: Now the phase boundary is determined by  Eq.~(\ref{eq:pb4}), which depends only on the cargo density on its left filament end. Using Eq.~(\ref{eq:boundary1}) and introducing the parameter,
\begin{equation}  
  C_{1} = \frac{Q}{2}(\frac{1}{2 W_+} - d)
\end{equation}
the equation of the phase boundary reads         
\begin{equation}         	
\alpha_{+} = \frac{C_{1}^{2} -2B_{1}C_{1}}{2 C_{1} W_{+} - Q}
\end{equation}
This phase boundary is a straight line parallel to the $\beta_+$ axis.

\item {\sl Phase boundary between (MC) and (LD-MC)}: The phase boundary is characterized by  Eq.~(\ref{eq:pb5}), which also depends only on the cargo density on its left filament end, as in the previous case. The equation of the phase boundary is,             
\begin{equation}         	
\alpha_{+} = \frac{C^{2} - 2B_{1}C}{2 C W_{+} -Q}
\end{equation}
This is also a straight line parallel to the $\beta_+$ axis.

\item {\sl Phase boundary between (HD) with (HD-MC)}: 
This boundary is characterized by Eq.~(\ref{eq:pb6}) and depends only on the cargo density on its right filament end. Using Eq.~(\ref{eq:boundary2})
\begin{equation}         	
\beta_{+} = (1-K_{+} -\alpha_{-}W_{+}) + \frac{V - Z_{4}^{2}}{2Z_{4}}
\end{equation}
This phase boundary is a straight line parallel to the $\alpha_+$ axis.

\item {\sl Phase boundary between (MC) with (HD-MC)}: Finally, the boundary line between these two phases obeys Eq.~(\ref{eq:pb7}) and and depends only on the cargo density on its right filament end. Using Eq.~(\ref{eq:boundary2}), we can explicitly write the expression for the phase boundary as a function of $\beta_+$ alone.
\begin{equation}         	
\beta_{+} = (1-K_{+} -\alpha_{-}W_{+}) + \frac{V - Z_{5}^{2}}{2Z_{5}}
\end{equation}
This phase boundary is also a straight line parallel to the $\alpha_+$ axis.
\end{enumerate}

\section{Deviations from the fast interconversion limit}

\begin{figure}
\begin{center}
\includegraphics[width=2.4in,angle=-90]{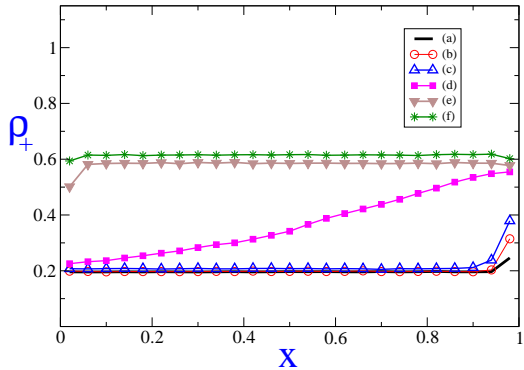}
\end{center}
\caption{(Color online)  Cargo density profile from MonteCarlo simulations with $N = 500$ as a function of the interconversion rate, $\tilde{q}$, for  $K_{+} = 0.25$, $\alpha_{+} / k_{t}= 0.2$ , $\beta_{+ }/k_{t}= 0.5$, $\alpha_{-} /k_{t}= 0.4 $, $\beta_{-} /k_{t}= 0.8$.  Langmuir kinetics is  switched off, $D_{\pm} =  E_{\pm} = 0$.  (a) $\tilde{q}\rightarrow \infty$(fast interconversion limit), (b) $\tilde{q} = 2$, (c) $\tilde{q}= 0.5$, (d) $ \tilde{q}= 0.25$, $k_{+} = 0.1$ (e) $ \tilde{q} = 0.1$ , (f) $\tilde{q} = 0.05$.}
\label{Fig10}
\end{figure}

Although we have focused in the limit where effective motor interconversion is the fastest process, we have carried out some Monte-Carlo studies to analyze the impact of such an assumption on collective motor dynamics. We compare the interconversion with the translation rate, $k_t$ through the parameter $\tilde{q} = k_{+} / k_t$, which is infinite in the fast interconversion limit. For example, for a given set of parameters where the density profile is homogeneous , as we decrease $\tilde{q}$ we move to a regime where the density profile is inhomogeneous and where its details depend separately on the two interconversion rates, $k_+$ and $k_-$. In this regime, the simplifying assumption of proportionality between the density of right and left moving cargoes no longer holds and a separate treatment for the two species will be required. Interestingly,  decreasing further $\tilde{q}$, we find typically a different homogeneous profile where the two cargo densities ( the left and right moving cargo densities) become proportional to each other again, where the profiles depend only on interconversion rates through $K_+$. Fig.~\ref{Fig10} shows a particular example where we start from an LD profile in the fast interconversion limit, for $\tilde{q}>0.5$, and move to an HD profile for $\tilde{q}<0.1$. An analysis of the deviations from the fast and slow interconversion limits and its impact in the nonequilibrium phase diagram remains an interesting challenge which requires a systematic study.

\begin{thebibliography}{}
\bibitem{example} For example, when $\rho_{R}(1) < \rho_{\mbox{lk}}$, then in the bulk,  the solution, $\rho_{R}(x)$ approaches the extremal solution  $W_{0}(-Y_{*}(x))$, introduced in Appendix A, where $Y_{*}(x)$ is the function defined by Eq.~(\ref{eq:Y}) evaluated for $x_{0} = 1$ and $\sigma(x_{0}) = -1$.
\bibitem{errata}  Although phase reentrance is  mentioned in Ref.~\cite{igna}, the location of the phase boundary between  (LD-HD) and (LD-MC-HD) is erroneously depicted.
\end{thebibliography}
\end{document}